\documentclass[%
 reprint,
 amsmath,amssymb,
 aps,
]{revtex4-2}

\usepackage{txfonts} 
\usepackage{braket} 
\usepackage{graphicx} 
\usepackage[unicode, colorlinks, citecolor=blue, linkcolor=red, urlcolor=blue]{hyperref}  
\usepackage{amsmath} 
\usepackage{comment} 
\usepackage{multirow} 
\usepackage{bm}

\begin{document}

\title{{Connection between the GKSL master equation and the Landauer formula}}
\author{Misa Nozaki}
\email{nozaki.misa@qst.go.jp}
\affiliation{%
 Institute for Quantum Life Science, National Institutes for Quantum Science and Technology (QST), Chiba 263-8555, Japan
}
\author{Takatoshi Fujita}%
 \email{fujita.takatoshi@qst.go.jp}
\affiliation{%
 Institute for Quantum Life Science, National Institutes for Quantum Science and Technology (QST), Chiba 263-8555, Japan
}%

%\author{Misa Nozaki$^1$\thanks{nozaki.misa@qst.go.jp} and Takatoshi Fujita$^1$\thanks{fujita.takatoshi@qst.go.jp} }
%\inst{$^1$Institute for Quantum Life Science, National Institutes for Quantum Science and Technology (QST), Chiba 263-8555, Japan} %\\

\begin{abstract}
%\abst{
We derive a current formula within the Gorini–Kossakowski–Sudarshan–Lindblad (GKSL) master equation formalism for a noninteracting system, and identify the conditions under which it reduces to the Landauer formula.
%
%and the assumption of an energy-independent Fermi distribution.
%In the derivation, we employ the solution of the Sylvester equation.
%}
\end{abstract}

\maketitle 

\begin{figure}[t]
\centering
\begin{tabular}{ll}
(a)&(b)\\
\includegraphics[bb=0 0 302 361, width=0.4\linewidth]{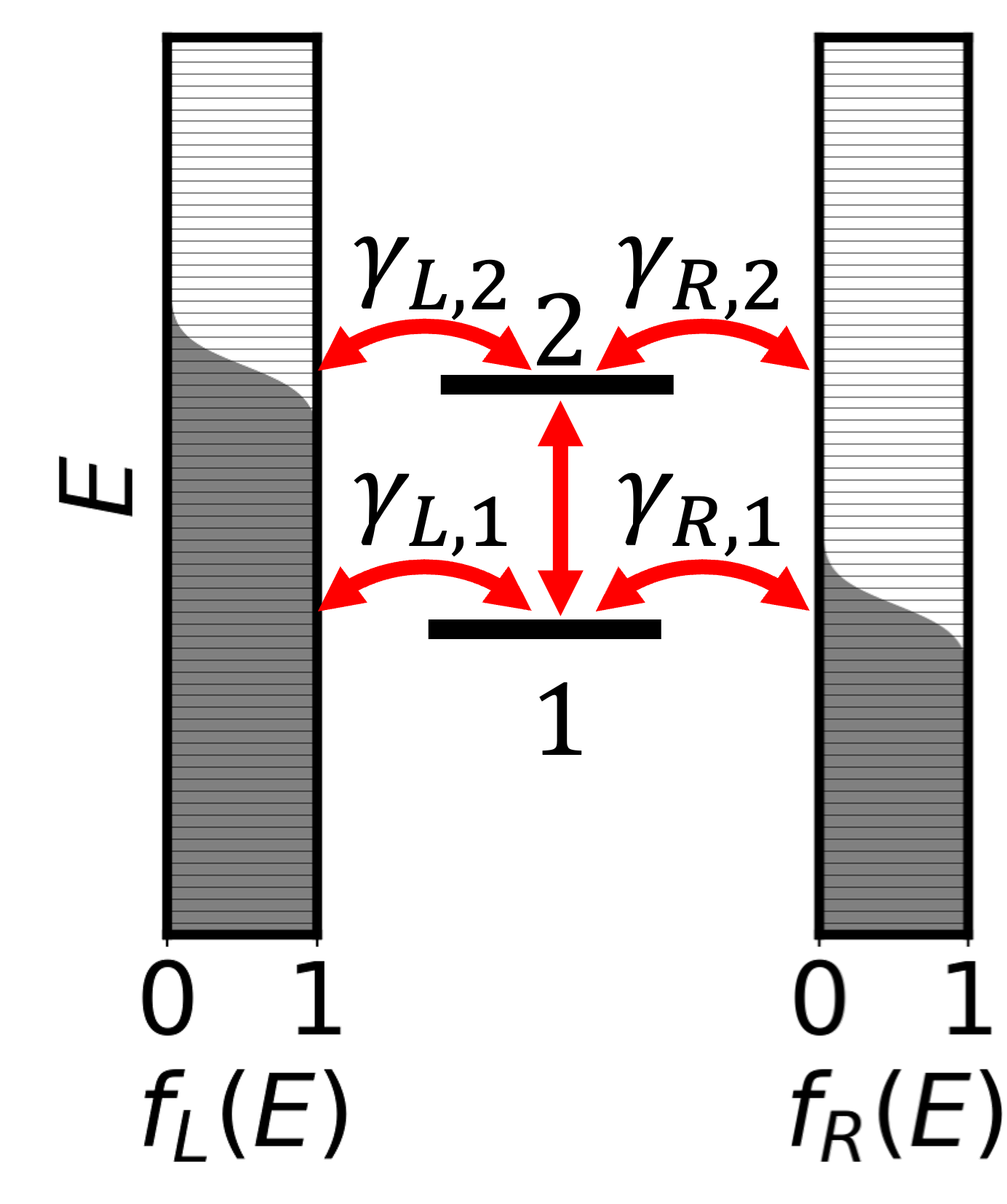}
&
\includegraphics[bb=0 0 302 361,width=0.4\linewidth]{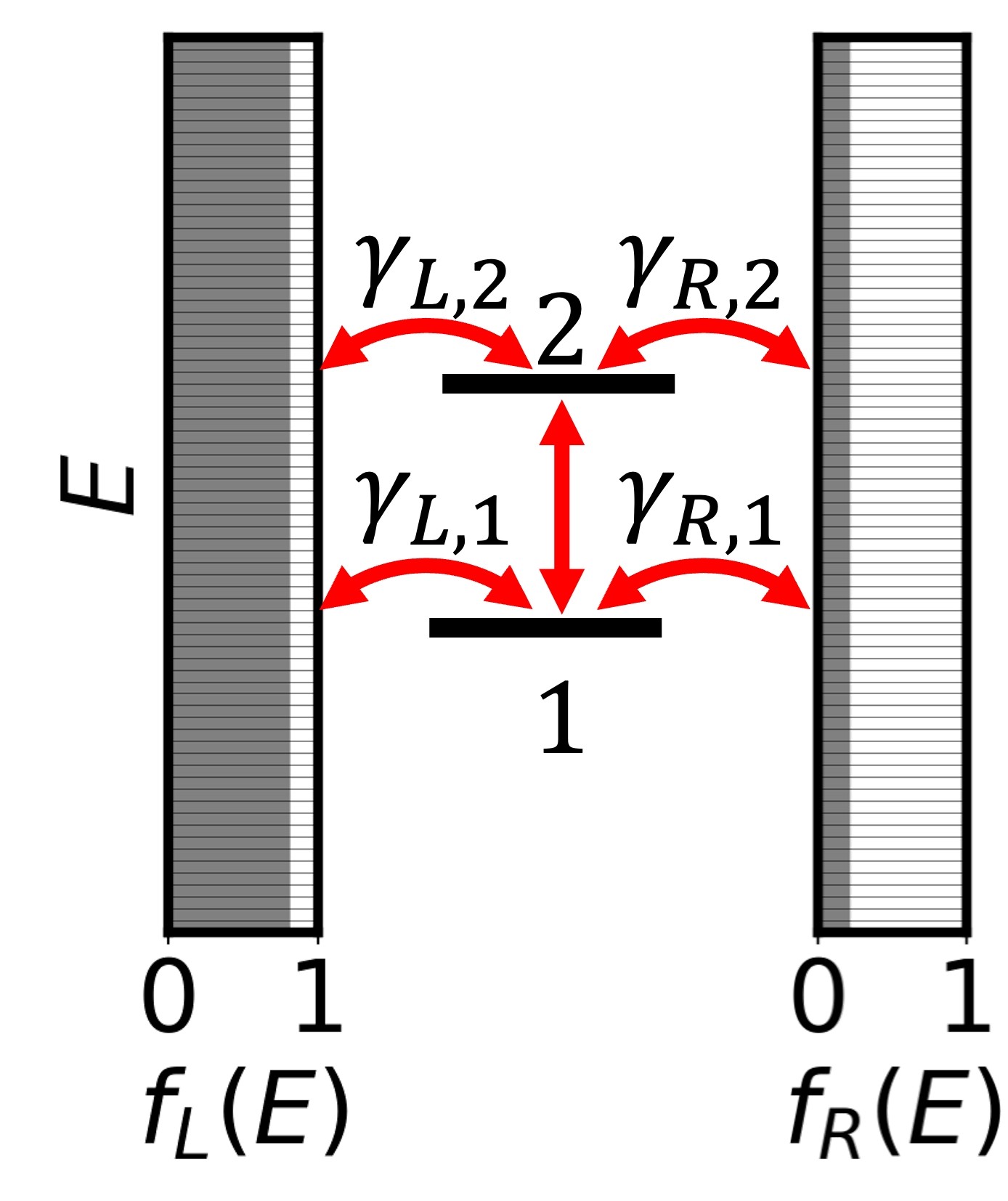}
\end{tabular}
\caption{Schematic illustration of the model considered in this work. The two rectangles represent left and right electrodes characterized by (a) energy-dependent general Fermi distribution functions and (b) energy-independent Fermi distribution functions. In the model, one electron orbital labeled $i$ is coupled to left/right electrodes with coupling strengths $\gamma_{L/R,i}$ ($i=1,2$), and coupling between different orbitals is also included.}
\label{fig:1}
\end{figure}

\section{Introduction}
Electron transport through mesoscopic systems, such as molecular junctions and quantum dots between two electrodes, has been extensively investigated using either the Green’s function approach \cite{stefanucci, Datta1995, MW} or the master equation approach \cite{Harbola2006, Timm2008, Nitzan2009, Timm2011, Oded2016, Guimaraes2016, Kolovsky_2024}. 
The choice between these approaches is largely dictated by the physical conditions and the purpose of the analysis. 
In particular, the Green’s function approach is well suited for describing coherent transport, whereas the master equation approach is commonly employed in weak coupling regimes, where dissipation, dephasing, and sequential tunneling processes play important roles.

Despite their widespread use, the connection between these two approaches is not yet fully understood. To clarify the connection, Jin {\it et~al}. derived an expression for the current for a specific system bath model within the framework of the Gorini–Kossakowski–Sudarshan–Lindblad (GKSL) master equation \cite{GKS, Lindblad} using a Keldysh formalism \cite{PhysRevB_Jin_2020}. 
The resulting formula coincides with the Meir–Wingreen formula \cite{MW}, and consequently with the Landauer formula for noninteracting electrons, in the limits of high temperature and large chemical potential under the wide-band limit approximation (WBLA). To gain further insight, analyses of electron transport within the Landauer and GKSL frameworks under more general conditions remain valuable.

%They mentioned the similarity of the obtained formula with the Landauer formula.
%, and demonstrated that the transport properties of a non-interacting fermionic chain connected to electrodes at both ends agree with those predicted by the Landauer formula in the high-temperature limit . %and from other view points are desired. % to investigate more general systems with, e.g., . 
%If the Landauer formula and the current within the GKSL formalism conincide for certain conditions should be further investigated. 

%The description based on Keldysh's action is not necessarily simple for all the readers. 
%Although two approach coincide in principle, the equivalence between two approach is unclear in the presence of approximations. 

%For elemental understanding, it is important to reveal the connection between these two approach in a certain approximation. To this end, Jin et al shows that Landauer and GKSL master equation is equivalent in a high temperature and high chemical potential limit via Keldysh formalism for fermionic chain \cite{PhysRevB_Jin_2020}, while Nitzan et al shows that their equivalence in weak coupling limit by assuming electrode explicitly\cite{Nitzan2009, Nitzan2020}. However, there is no work that derive general condition for equivalence between two approach.

In this work, we consider a non-interacting electron system connected to left(L) and right(R) electrodes characterized by Fermi distribution functions, within the framework of the GKSL master equation. We derive a current formula for this system and clarify the conditions under which it coincides with the Landauer formula.

\section{Theoretical Analysis}
\subsection{Model setting}
In the following, we consider the electronic Hamiltonian of the system, given by
\begin{align}
 \hat{H} &= \sum_{ij} h_{ij} \hat{c}_i^\dagger \hat{c}_j
 %+\sum_{ijkl} v_{ijkl} \hat{c}_i^\dagger \hat{c}_j^\dagger {\hat c}_k {\hat c}_l ,
 \label{eq:1}
\end{align}
where $\hat{c}_{i}^{\dagger}$ and $\hat{c}_{i}$ are the electron creation and annihilation operators for orbital $i$ ($i = 1, 2, \ldots, M$), and $h_{ij}$ are the matrix elements of the Hamiltonian ${\bm h}$. Since the GKSL master equation requires the Markov approximation, we assume the WBLA and consider a regime in which the coupling between each orbital $i$ and the electrodes is sufficiently weak, so that Fermi distribution can be approximated as energy independent within the energy window set by the coupling strength. This situation is schematically illustrated in Fig.~\ref{fig:1}.

\subsection{Landauer formula}
Within the WBLA, the Landauer formula can be written as
\begin{align}
&I=\frac{q}{h}\int_{-\infty}^{\infty} d\epsilon \mathrm{Tr}\left[ {\bm \gamma}_L {\bm G}^r(\epsilon) {\bm \gamma}_R {\bm G}^a(\epsilon) \right] \left(f_L(\epsilon)-f_R(\epsilon)\right)\label{eq:2},\\
&{\bm G}^{r}(\epsilon)=\left(\epsilon-{\bm h}+ \frac{i({\bm \gamma}_L+{\bm \gamma}_R)}{2} + i\eta \right)^{-1}, \label{eq:3}\\
&{\bm G}^{a}(\epsilon)=\left(\epsilon-{\bm h}-\frac{i({\bm \gamma}_L+{\bm \gamma}_R)}{2} - i\eta \right)^{-1}. \label{eq:4}
\end{align}
Here, $q$ is the electron charge, $h$ is the Planck constant, $\epsilon$ is energy, and $\eta$ is an infinitesimal positive constant. 
The matrices ${\bm \gamma}_{L/R}$ denote the level broadening due to the left/right electrodes. ${\bm G}^{r(a)}$ are the retarded (advanced) Green’s functions of the scattering region, including the effects of the electrodes through the self-energies, which reduce to $\mp i\,{\bm \gamma}_{L/R}/2$ within the WBLA. 
The functions $f_{L/R}(\epsilon)$ are the Fermi distribution functions of the left/right electrodes.

\subsection{GKSL master equation formalism}
Within the GKSL master equation formalism, time evolution of the density matrix under the assumption mentioned above can be described by the following equation:
\begin{eqnarray}
\frac{d\rho}{dt}&=&-\frac{i}{\hbar}[{\hat H},\rho] + \mathcal{L}[\rho],
\label{eq:5}\\
\mathcal{L}[\rho]&=&\sum_{a=R,L} \sum_{i}
\left\{
\alpha_{a,i} 
\left[ {\hat c}_i^{\dagger} \rho {\hat c}_i - 
\frac{1}{2}\left\{ {\hat c}_i{\hat c}_i^{\dagger} , \rho \right\} 
\right] \right . \nonumber \\
&+& \left.\beta_{a,i}
\left[ {\hat c}_i \rho {\hat c}_i^{\dagger} - 
\frac{1}{2}\left\{ {\hat c}_i^\dagger {\hat c}_i, \rho \right\} 
\right ]
\right\},
\label{eq:6}
\end{eqnarray}
where $\alpha_{a,i}=\gamma_{a,i}f_{a,i}/\hbar$  and $
\beta_{a,i}=\gamma_{a,i}(1-f_{a,i})/\hbar$. $\gamma_{L(R),i}$ represent coupling strength between electrode $L(R)$ and orbital $i$, and $f_{a,i}$ represent the Fermi distribution with energy $h_{ii}$ ($f_{a,i}=f_a(h_{ii})$). 
The electric current is related to the time evolution of the total number of electrons in the system, $N = \textrm{Tr}\left[\sum_{i} {\hat c}_i^\dagger {\hat c}_i \rho \right]$. By using Eq. \eqref{eq:5}, 
\begin{align}
\frac{dN}{dt}&=%\frac{d}{dt}\Tr\left[\sum_{i} {\hat c}_i^\dagger {\hat c}_i \rho \right]
\frac{1}{q}\left( I_L+I_R \right),
\label{eq:7}\\
I_a&= q\sum_i \left\{ 
\alpha_{i,a}\operatorname{Tr} \left[ \rho {\hat c}_i {\hat c}_i^\dagger \right]       -\beta_{i,a}\operatorname{Tr}\left[ \rho {\hat c}_i^{\dagger} {\hat c}_i \right] 
\right \} \nonumber \\
&= \frac{q}{\hbar}\operatorname{Tr}\left [ \bm{\gamma}_a ( {\bm f}_a-\bm{n})\right ],\quad (a=L,R). \label{eq:8}
\end{align}
Here, $\bm{n}$ is the single electron density matrix with elements of $n_{ij}=\operatorname{Tr} \left[{\hat c}_i^\dagger {\hat c}_j \rho \right]$. ${\bm \gamma}_{a}$ and ${\bm f}_{a}$ are diagonal matrices with elements of  $\gamma_{a,ij}=\gamma_{a,i}\delta_{ij}$ and $f_{a,ij}=f_{a,i}\delta_{ij}$, respectively.
$I_{L(R)}$ corresponds to the net current injected from left (right) electrode.

Within the steady state condition, the single electron density matrix $n_{ij}$ no longer depends on time. 
%Here, we consider steady state current in Lindblad master equation [Eq. \eqref{eq:8}] with one-electron system described as 
%\begin{eqnarray}
%{\hat H}=\sum_{i,j} H_{ij} {\hat c}_i^\dagger {\hat c}_j \label{eq:9}.
%\end{eqnarray}
%From Eq. \eqref{eq:5}, 
%The single electron density matrix satisfy $\frac{d\operatorname{Tr}\left[{\hat c}_i^\dagger {\hat c}_j \rho \right]}{dt}=0$ in steady state, and then $n_{ij}=\operatorname{Tr} \left[{\hat c}_i^\dagger {\hat c}_j \rho \right]$ satisfy following equation:
Thus, using $n_{ij}$ and Eq. \eqref{eq:5}, we obtain
\begin{eqnarray} 
\frac{d{\bm n}}{dt} &=&
-i[{\bm h},{\bm n}]-\frac{1}{2}\left\{{\bm \gamma}_L+{\bm \gamma}_R, {\bm n}\right \}
\nonumber \\
&+&({\bm \gamma}_L {\bm f}_L + {\bm \gamma}_R {\bm f}_R)={\bm 0}.
\label{eq:9}
\end{eqnarray}
Eq. \eqref{eq:9} reduces to Sylvester equation:
%can be rewritten as,
\begin{align}
&\left (-{\bm h}+\frac{i({\bm \gamma}_L+{\bm \gamma}_R)}{2}\right){\bm n} - {\bm n} \left ( -{\bm h}- \frac{i({\bm \gamma}_L+{\bm \gamma}_R)}{2}\right )\nonumber \\
&=i({\bm \gamma}_L {\bm f}_L + {\bm \gamma}_R {\bm f}_R).
\label{eq:10}
\end{align}
%Eq. \eqref{eq:11} has the same form of sylvester equation $AX-XB=Y$. By using the general solution $X=\frac{1}{2\pi i} \int_\Gamma (A-\zeta)^{-1}Y(B-\zeta)^{-1}d\zeta$ [theorem VII.2.4 in Ref. \citen{bhatia1997matrix}], we obtain 
The Sylvester equation, ${\bm A}{\bm X}-{\bm X}{\bm B}={\bm Y}$, for ${\bm X}$ has a general solution of ${\bm X}=\frac{1}{2\pi i} \int_\Gamma ({\bm A}-\zeta)^{-1}{\bm Y}({\bm B}-\zeta)^{-1}d\zeta$ (See for details Theorem VII.2.4 in Ref. \cite{bhatia1997matrix}). 
By using the formula, we obtain 
\begin{align}
&{\bm n}=\frac{1}{2\pi}\int_{-\infty}^{\infty} {\bm G}^r(\epsilon)({\bm \gamma}_L {\bm f}_L + {\bm \gamma}_R {\bm f}_R){\bm G}^a(\epsilon)d\epsilon, 
\label{eq:11}\\
&{\bm G}^{r}(\epsilon)=\left(\epsilon-{\bm h}+ \frac{i({\bm \gamma}_L+{\bm \gamma}_R)}{2} +i\eta \right)^{-1}, \label{eq:12}\\
&{\bm G}^{a}(\epsilon)=\left(\epsilon-{\bm h}-\frac{i({\bm \gamma}_L+{\bm \gamma}_R)}{2} -i\eta \right)^{-1}. \label{eq:13}
\end{align} 
Here, we add innfinitesimal $\eta$ following the theorem. 
%Since the diagonal elements of $\bm{n}$ are real, the %correponding elements in Eq. \eqref{eq:12} are invariant %under complex conjugation:
%\begin{eqnarray}
%n_{ii}=\frac{1}{2\pi}\int_{-\infty}^{\infty} 
%\left[ G^a(\epsilon)(\gamma_L f_L + \gamma_R %f_R)G^r(\epsilon) \right]_{ii}
%d\epsilon
%\label{eq:12_cc}
%\end{eqnarray} 
Substituting Eqs. \eqref{eq:11}-\eqref{eq:13} into Eq. \eqref{eq:8}, the current from left electrode to the system can be rewritten as 
\begin{align}
I_L&=\frac{q}{\hbar}\operatorname{Tr}\left[ {\bm \gamma}_L({\bm f}_L-{\bm n})\right] \nonumber \\
&=\frac{q}{\hbar} \sum_{i} \gamma_{L,i} (f_{L,i}-n_{ii} ) \nonumber \\
&=
\frac{q}{h}\sum_{i} \gamma_{L,i} f_{L,i}
\sum_{m}
\int_{-\infty}^{\infty} d\epsilon G^a_{im}(\epsilon)(\gamma_{L,m}+\gamma_{R,m})G^a_{mi}(\epsilon)\nonumber \\
&-\frac{q}{h}\sum_{i} \gamma_{L,i} 
\sum_{m}
\int_{-\infty}^{\infty} d\epsilon G^r_{im}(\epsilon)(\gamma_{L,m}f_{L,m} - \gamma_{R,m}f_{R,m})G^a_{mi}(\epsilon)
\nonumber \\
&=\frac{q}{h}\int_{-\infty}^{\infty} d\epsilon  \operatorname{Tr}[{\bm f}_L{\bm \gamma}_L {\bm G}^a(\epsilon) {\bm \gamma}_{R} {\bm G}^r(\epsilon)-{\bm f}_R{\bm \gamma}_R {\bm G}^a(\epsilon) {\bm \gamma}_{L} {\bm G}^r(\epsilon)].\label{eq:14}
\end{align}
Thus, the current within the GKSL formalism can be expressed in terms of Green’s functions. For the transformation from the second to the third line of Eq. \eqref{eq:14}, we use
\begin{eqnarray}
\sum_m\frac{1}{2\pi}\int_{-\infty}^{\infty} d\epsilon G^{a}_{im}(\epsilon) (\gamma_{L,m}+\gamma_{R,m})G^{r}_{mi}(\epsilon) 
= 1\label{eq:15}
\end{eqnarray}
which follows from the relation 
$i\left( {\bm G}^r(\epsilon)-{\bm G}^a(\epsilon)\right)={\bm G}^r(\epsilon)({\bm \gamma}_L +{\bm \gamma}_R){\bm G}^a(\epsilon)={\bm G}^a(\epsilon)({\bm \gamma}_L +{\bm \gamma}_R){\bm G}^r(\epsilon)$ together with the spectral sum rule. 

While Eqs. \eqref{eq:2} and \eqref{eq:14} generally differ, they coincide under specific conditions. First, when the system is described by a diagonal Hamiltonian matrix—i.e., when it consists of independent multichannel modes—Eq. \eqref{eq:14} can be evaluated as
\begin{eqnarray}
I_L=\frac{q}{h}\sum_i (f_{L,i}-f_{R,i})\gamma_{L,i}\gamma_{R,i}\frac{2\pi}{\gamma_{L,i}+\gamma_{R,i}}. 
\end{eqnarray}
This expression coincides with the Landauer formula in the weak electrode–system coupling limit. Furthermore, the two expressions also agree when the distribution functions $f_a(E)$ can be regarded as energy-independent constants $f_a$ $(a = L, R)$ within the energy range of interest, as schematically illustrated in Fig.~\ref{fig:1}(b). In this case, both Eq. \eqref{eq:2} and Eq. \eqref{eq:14} reduce to the same simplified form:
\begin{eqnarray}
\frac{q}{h}\int_{-\infty}^{\infty} d\epsilon \operatorname{Tr}[{\bm\gamma}_L {\bm G}^r(\epsilon) {\bm\gamma}_{R} {\bm G}^a(\epsilon)](f_L-f_R). \label{eq:16}
\end{eqnarray}
Note that $f_{L}(\epsilon)=1$ and $f_{R}(\epsilon)=0$ are justified when all single-particle orbital energies $\epsilon_i$ lie between the chemical potentials of the left ($\mu_L$) and right ($\mu_R$) electrodes ($\mu_R < \epsilon_i < \mu_L$) and the temperature is sufficiently low compared to the bias window ($k_B T \ll \mu_L - \mu_R$).

\section{Conclusion}
To conclude, we derived the electric current through a system between two electrodes within the GKSL master equation formalism, which is directly comparable to the Landauer's expression. 
The derivation was via the general solution to the Sylvester's equation. 
We clarified that the two approaches become equivalent when the energy dependence of the Fermi distributions can be neglected within the relevant energy range.

\begin{acknowledgments}
This work was supported by Grant-in-Aid for Transformative Research Areas ``Materials Science of Meso-Hierarchy'' (No.JP23H04879). 
\end{acknowledgments}

\bibliography{ref}

\end{document}